\documentclass[english]{article}
\usepackage[T1]{fontenc}
\usepackage[latin9]{inputenc}
\usepackage[active]{srcltx}
\usepackage{color}
\usepackage{float}
\usepackage{amsmath}
\usepackage{amssymb}
\usepackage{graphicx}
\usepackage{setspace}
\usepackage{enumitem}
\usepackage[font={footnotesize,it}]{caption}
\usepackage[font={footnotesize,it}]{subcaption}

\makeatletter

\usepackage{spconf}\usepackage{epsfig}

\address{\small $^{1}$Dpt. of Computer, Control, and Management Eng., University of Rome ``La Sapienza", Roma, Italy.\\ \small
$^2$Dpt. of Electrical Eng., State University of New York (SUNY) at Buffalo, Buffalo, NY 14260, USA.
\vspace{-0.2cm}}

\name{Francisco Facchinei$^{1}$, Simone Sagratella$^{1}$, Gesualdo Scutari$^{2}$\vspace{-0.4cm}\thanks{\noindent The work of Scutari was supported by the USA National Science Foundation under Grants CMS 1218717 and  CAREER Award No. 1254739}}

\renewcommand{\Re}{\mathbb{R}}

\newcommand{\z}{\mathbf{z}}
\newcommand{\y}{\mathbf{y}}
\newcommand{\x}{\mathbf{x}}

\newcommand{\A}{\mathbf{A}}
\newcommand{\bv}{\mathbf{b}}
\newcommand{\Q}{\mathbf{Q}}
\newtheorem{theorem}{Theorem}
\newtheorem{proposition}[theorem]{Proposition}

\newtheorem{lemma}[theorem]{Lemma}

\newcounter{algo}
\newenvironment{algo}[2]{\refstepcounter{algo}\label{#2}   \begin{center}
\begin{minipage}{0.49\textwidth}   \hrule\smallskip
\textbf{Algorithm \thealgo: #1}
\par\smallskip\hrule\smallskip\ignorespaces}{\par\smallskip\hrule
\end{minipage}
\end{center}
}

\usepackage{babel}

\makeatother

\usepackage{babel}
\begin{document}
\ninept

\title{\vspace{-1.3cm}\textbf{
Flexible Parallel    Algorithms for   Big Data Optimization
\ninept
}}
\maketitle
\begin{abstract}
We propose  a  decomposition framework for the parallel optimization
of the sum of a differentiable function and a (block) separable
nonsmooth, convex one. The latter term is typically used to enforce structure in the solution
as, for example, in Lasso problems.
Our framework is very flexible and includes both fully parallel Jacobi schemes
and Gauss-Seidel (Southwell-type) ones, as well as virtually all possibilities in between (e.g.,  gradient- or Newton-type
methods) with only a subset of  variables  updated at each iteration. Our theoretical convergence results improve on existing ones, and
  numerical results show  that the new method compares favorably to existing
algorithms.
\end{abstract}

\noindent
\begin{keywords}   Parallel  optimization,  Jacobi method, Lasso, Sparse solution. \end{keywords}
\vspace{-0.2cm}

\section{Introduction\vspace{-0.1cm}}

\label{sec:Introduction}

The minimization of the sum of a smooth  function, $F$, and of a nonsmooth
(separable) convex one, $G$,
\begin{equation}\label{eq:problem 1}
\min_{\x\in X} V(\x) \triangleq F(\x) + G(\x),
\end{equation}
is an ubiquitous problem that arises in many fields of
engineering, so diverse as compressed sensing, basis pursuit denoising,
sensor networks, neuroelectromagnetic imaging, machine learning,
data mining, sparse logistic regression,  genomics,  metereology, tensor factorization and completion,
geophysics, and radio astronomy.
Usually the nonsmooth term is used to promote sparsity of the optimal solution,
that often corresponds to a parsimonious representation of some phenomenon at hand.
Many of the mentioned applications
can give rise to extremely large problems so that standard optimization techniques are hardly applicable.
And indeed, recent years have witnessed a flurry of research activity  aimed at developing
solution methods that are simple (for example based solely on matrix/vector multiplications) but yet capable to
converge
to a good approximate solution in reasonable time. It is hardly possible here to even summarize   the huge amount
of work  done in this field; we refer the reader to the recent works
 \cite{bach2011optimization,bradley2011parallel,buhlmann2011statistics,byrd2013inexact,fountoulakis2013second,necoara2013efficient,nesterov2012gradient,
 nesterov2012efficiency,qin2010efficient,rakotomamonjy2011surveying,razaviyayn2013unified,richtarik2012iteration,
 richtarik2012parallel,Sra-Nowozin-Wright_book11,tseng2009coordinate,xu2012block,yin2013parallel,
 yuan2010comparison,wright2012accelerated}
 as entry points to the literature.

It is clear however that if one wants to solve really large problems,
parallel methods  exploiting the computational  power of multi-core processors have to be employed.
It is then surprising that while serial solutions methods for Problem \eqref{eq:problem 1} have been widely investigated,
the analysis of  parallel algorithms suitable to large-scale implementations  lags behind.
Gradient-type methods can of course be easily parallelized. However, beyond that,
we are only aware of very few papers, all very recent,
that deal with parallel solution methods \cite{bradley2011parallel,necoara2013efficient,  richtarik2012parallel,yin2013parallel}.
These papers analyze  both randomized and deterministic block Coordinate Descent Methods (CDMs) that, essentially,
 are still (regularized) gradient-based methods. One advantage of the analyses in    \cite{bradley2011parallel,necoara2013efficient,  richtarik2012parallel,yin2013parallel}
is that they provide an interesting  (global) rate of convergence. On the other hand they apply only to convex problems and
are not  flexible enough to include, among other things,  very natural Jacobi-type methods (where at each iteration
a partial minimization  of the original function is performed with respect to a block variable while all other variables are kept fixed) and  the possibility to deal with a nonconvex $F$.

In this paper, building on the approach proposed in
\cite{scutari_facchinei_et_al_icassp13,scutari_facchinei_et_al_tsp13}, we present a  broad, deterministic  algorithmic framework for the solution of Problem \eqref{eq:problem 1} with the following novel features:
i) it is parallel, with a degree of parallelism that can be chosen by the user and that can go from a complete parallelism  (each variable is updated
in parallel to all the others) to the sequential (one variable only is updated at each iteration); ii) it can tackle a nonconvex $F$; iii)
it is very flexible and  includes, among others,  updates based on gradient- or Newton-type methods;  and iv) it easily allows for inexact
solutions.
Our framework allows us to define different schemes, \emph{all converging under the same conditions}, that can accommodate different problem features and  algorithmic requirements.
 Even in the most studied case in which $F$ is convex and $G(\x) \equiv 0$ our results compare favourably to existing ones and the numerical results
 show our
 approach to be very promising.
 \vspace{-0.1cm}

\section{Problem Definition\vspace{-0.1cm}}

We consider Problem \eqref{eq:problem 1},
where the feasible set $X=X_1\times \cdots \times  X_N$ is a cartesian product of lower dimensional convex sets $X_i\subseteq \Re^{n_i}$, and $\mathbf x\in \Re^n$ is partitioned accordingly to $\mathbf x = (\mathbf  x_1, \ldots, \mathbf x_N)$, with each $\mathbf  x_i \in \Re^{n_i}$. $F$ is  smooth (and not necessarily convex)
and $G$  is convex and possibly nondifferentiable, with  $ G(\x) = {\scriptstyle \sum_{i=1}^N} g_i(\x_i)$ with $\x_i \in X_i$.
This format is very general and includes problems of great interest.
Below we list some  instances of Problem \eqref{eq:problem 1}.


\noindent $\bullet$
$G(\x)=0$; in this case the problem reduces to the minimization of a smooth, possibly nonconvex problem with convex constraints.

\noindent $\bullet$
$ F(\x) = \|\A\x -\bv\|^2 $ and $G(\x)= c \|\x\|_1$, $X=\Re^n$, with $\A\in\Re^{m \times n}$, $\bv \in \Re^m$,  and $c\in \Re_{++}$    given constants; this is the very famous and much studied Lasso problem \cite{tibshirani1996regression}.

\noindent $\bullet$
 $ F(\x) = \|\A\x -\bv\|^2 $ and $G(\x)= c \sum_{i=1}^N\|\x_i\|_2$, $X=\Re^n$, with $\A\in\Re^{m \times n}$, $\bv \in \Re^m$,  and $c\in \Re_{++}$
given constants; this is the group Lasso problem \cite{yuan2006model}.

\noindent $\bullet$
$ F(\x) = \sum_{j=1}^m \log(1 + e^{-a_i \y_i^T\x }) $ and $G(\x)= c \|\x\|_1$ (or $G(\x)= c \sum_{i=1}^N\|\x_i\|_2$), with $\y_i\in \Re^n$, $a_i\in \Re$, and $c\in \Re_{++}$  given
constants; this is the sparse logistic regression problem \cite{shevade2003simple,meier2008group}.

\noindent $\bullet$
$F(\x) = \sum_{j=1}^m\max\{0, 1- a_i \y_i^T\x\}^2$ and $G(\x)= c \|\x\|_1$, with $a_i\in \{-1,1\}$,  $\y_i\in \Re^n$, and  $c\in \Re_{++}$ given; this is
the $\ell_1$-regularized
$\ell_2$-loss Support Vector Machine problem, see e.g. \cite{yuan2010comparison}.

\noindent $\bullet$ Other problems that can be cast in the form \eqref{eq:problem 1} include the Nuclear Norm Minimization problem,  the Robust Principal Component Analysis problem, the Sparse Inverse Covariance Selection problem, the Nonnegative Matrix (or Tensor) Factorization problem, see e.g. \cite{xu2012block,goldfarb2012fast} and references therein.


\smallskip

\noindent
Given \eqref{eq:problem 1}, we make the following standard, blanket assumptions:

\begin{description}[topsep=-2.0pt,itemsep=-2.0pt]
\item[\rm  (A1)]  Each $X_i$ is nonempty, closed, and convex;
\item[\rm  (A2)] $F$ is $C^1$ on an open set containing $X$;
\item[\rm  (A3)]  $\nabla F$ is   Lipschitz continuous
on $X$ with constant $L_{F}$; 
\item[\rm  (A4)] $G(\x) = \sum_{i=i}^N g_i(\x_i)$, with all $g_i$ continuous and convex on $X_i$;
\item[\rm  (A5)] $V$ is coercive.
\end{description}
\vspace{-0.2cm}

\section{Main Results\label{sec:Main Results}\vspace{-0.2cm}}

We want to develop  {\em parallel} solution methods for Problem \eqref{eq:problem 1} whereby operations can be carried out on some or  (possibly) all   (block) variables $\x_i$ at
the \emph{same} time. The most natural   parallel (Jacobi-type) method one can think of is  updating  \emph{all} blocks simultaneously: given $\mathbf{x}^k$, each (block) variable $\mathbf{x}^{k+1}_i$ is computed as the solution of
$\min_{\mathbf{x}_i}$ $[F(\mathbf{x}_i, \mathbf{x}_{-i}^k) + g_i(\mathbf{x}_i)] $  (where $\x_{-i}$ denotes the vector obtained from $\x$ by deleting the block $\x_i$).
Unfortunately this method converges only under very restrictive conditions  \cite{Bertsekas_Book-Parallel-Comp} that are seldom verified in practice. To cope with this issue  we introduce some ``memory"
and set the new point to be a convex combination of $\mathbf{x}^k$ and the solutions of $\min_{\mathbf{x}_i} [F(\mathbf{x}_i, \mathbf{x}^k_{-i}) + g_i(\mathbf{x}_i)] $.
However our framework has many additional features, as discussed next.

\noindent \textbf{Approximating $F$}: Solving each  $\min_{\mathbf{x}_i} [F(\mathbf{x}_i, \mathbf{x}^k_{-i}) + g_i(\mathbf{x}_i)] $ may be too costly or difficult in
some
situations. One may then
prefer to approximate this  problem, in some suitable sense,  in order to facilitate the task of computing the new iteration.
To this end,
we assume that for all  $i\in {\cal N} \triangleq \{1, \dots, N\}$ we
can define  a function  $P_{i} (\mathbf{z};\mathbf{w}) :  X_i \times  X \to \Re$   having the following properties (we denote by $\nabla P_{i}$
the partial gradient of $P_i$ with respect to  $\mathbf{z}$):
\begin{description}[topsep=-1.0pt,itemsep=-2.0pt]
\item[\rm  (P1)]
 $P_{i} (\mathbf{\bullet}; \mathbf{w})$ is convex and continuously differentiable
on $ X_i$ for all $\mathbf{w}\in X$;
\item[\rm  (P2)]  $\nabla P_{i} (\mathbf{x}_i;\mathbf{x}) = \nabla_{{\mathbf{x}}_i} F(\mathbf{x})$ for all $\mathbf{x} \in  X$;
\item[\rm  (P3)]  $\nabla P_{i} (\mathbf{z};\mathbf{\bullet})$ is Lipschitz continuous
on $ X$ 
 for all $\mathbf{z} \in  X_i$.
\end{description}
\smallskip

Such a  function  $P_i$ should be regarded as a (simple) convex approximation of $F$ at the point $\mathbf{x}$ with respect to the block of variables $\mathbf{x}_i$, that
preserves the first order properties of $F$
with respect to  $\mathbf{x}_i$.
Based on this approximation we can define at any point $\x^k\in X$  a {\em regularized} approximation  $\widetilde{h}_i (\mathbf{x}_{i};\mathbf{x}^k)$ of $V$ with respect
to $\mathbf{x}_i$ where $F$ is replaced  by $P_i$ while  the nondifferentiable term is preserved,  and a quadratic
regularization is added to make the overall approximation strongly convex. More formally, we have
\[
\widetilde{h}_i (\mathbf{x}_{i}; \mathbf{x}^k) \! \triangleq \!\underbrace{  P_{i}(\mathbf{x}_{i}; \mathbf{x}^k)\!
 +\!\dfrac{{\tau}_{i}}{2} \! \left(\mathbf{x}_{i}- \mathbf{x}_{i}^k\right)^{T}\!\!\!\!\Q_{i}( \mathbf{x}^k)\!\left(\mathbf{x}_{i}- \mathbf{x}^k_{i}\right)}_{\triangleq h_i(\mathbf{x}_{i}; \mathbf{x}^k) }
 \!+ g_{i}(\mathbf{x}_{i}),\label{eq:convex_approx_of_fi_on_Ci}\vspace{-0.2cm}
\]
where $\Q_{i}(\mathbf{x}^k)$ is an $n_{i}\times n_{i}$
 positive definite matrix (possibly dependent on $\mathbf{x}^k)$, satisfying the following conditions. \smallskip
 \begin{description}[leftmargin=*,topsep=-2.0pt,itemsep=-2.0pt]
\item[\rm  (A6)] All matrices  $\Q_{i}(\mathbf{x}^k)$ are uniformly positive definite with a common positive definiteness constant $q > 0$; furthermore,  $\Q_{i}(\bullet)$ is Lipschitz continuous on $X$.\smallskip
\end{description}

 Note that in most cases (and in all our numerical  experiments) the $\Q_{i}$ are constant and equal to the identity matrix, so that (A6) is automatically satisfied. 
Associated with each $i$ and point $ \mathbf{x}^k \in X$ we can define the following optimal solution map:\vspace{-0.1cm}
\begin{equation}\label{eq:decoupled_problem_i}
\widehat{\mathbf{x}}_{i}(\mathbf{x}^k,\tau_{i})\triangleq\underset{\mathbf{x}_{i}\in X_{i}}{\mbox{argmin}\,}\tilde{h}_{i}(\mathbf{x}_{i};\mathbf{x}^{k}).\vspace{-0.1cm}
\end{equation}
 Note that $\widehat{\mathbf{x}}_{i}(\mathbf{x}^{k},\tau_{i})$
is always well-defined, since the optimization problem in (\ref{eq:decoupled_problem_i})
is strongly convex. Given (\ref{eq:decoupled_problem_i}),
we can then introduce
\[
X\ni\mathbf{y}\mapsto\widehat{\x}(\mathbf{y},\boldsymbol{{\tau}})\triangleq\left(\widehat{\x}_{i}(\y,\tau_{i})\right)_{i=1}^{N}.
\]
The algorithm we are about to  described is based on the computation of  $\widehat{\x}$. Therefore the approximating functions $P_i$ should
lead to as easily computable functions $\widehat{\x}$ as possible. An appropriate choice depends on the problem at hand and   on computational
requirements. We discuss some  possible choice for the  approximations $P_i$ after introducing the main algorithm (Algorithm \ref{alg:general}).

\noindent \textbf{Inexact solutions:}  In many situations (especially in the case of large-scale problems), it can be useful to
further reduce the computational effort needed to solve the subproblems in \eqref{eq:decoupled_problem_i} by allowing \emph{inexact} computations $\mathbf{z}^k$ of $\widehat{\mathbf{x}}_{i}(\mathbf{x}^k,\tau_{i})$, i.e., $\|\mathbf{z}_{i}^{k}-\widehat{\mathbf{x}}_i\left(\mathbf{x}^{k},\boldsymbol{{\tau}}\right)\|\leq\varepsilon_{i}^{k}$, where $\varepsilon_{i}^{k}$ measures the   accuracy in computing the solution.

\noindent \textbf{Updating  only some blocks:} Another important feature of our algorithmic framework  is the  possibility of updating only some of the variables at each iteration.
Essentially we prove convergence assuming that at each iteration only a subset of
the variables is updated under the condition that this subset contains at least one (block) component
which is within a factor $\rho \in (0,1]$ from being  ``sufficiently far'' from optimality, in the sense explained next.
First of all $\x^k_i$ is optimal for $\tilde{h}_{i}(\x_{i};\x^k)$ if and only if
$\widehat{\x}_{i}(\x^k,\tau_{i})=\x^k_i$. Ideally we would  like then to select the indices to update according to the optimality measure $\|\widehat{\x}_{i}(\x^k,
\tau_{i})-\x^k_i\|$; but in some situations this    could be  computationally  too expensive.  In order to be able to develop alternative choices, based
on the same idea, we   suppose  one can compute an {\em error bound}, i.e., a function $E_i(\x)$ such that\vspace{-0.1cm}
\begin{equation}\label{eq:error bound}
\underbar s_i\|\widehat{\x}_{i}(\x^k,\tau_{i})-\x^k_i\| \le E_i(\x^k) \leq \bar s_i\|\widehat{\x}_{i}(\x^k,\tau_{i})-\x^k_i\|,\vspace{-0.1cm}
\end{equation}
for some $0< \underbar s_i \le \bar s_i$. Of course we can always set  $ E_i(\x^k) = \|\widehat{\x}_{i}(\x^k,\tau_{i})-\x^k_i\| $, but other choices are also possible; we discuss
some of them after introducing the algorithm.

We are now ready to formally introduce our algorithm, Algorithm 1,  that enjoys all the features discussed above. Its convergence properties are given in Theorem \ref{Theorem_convergence_inexact_Jacobi}, whose proof is omitted because of space limitation, see \cite{FacchineiSagratellaScutariMPsub13}.

\vspace{-0.4cm}

\begin{algo}{\textbf{Inexact Parallel Algorithm}} S$\textbf{Data}:$ $\{\varepsilon_{i}^{k}\}$
for $i\in\mathcal{N}$, $\boldsymbol{{\tau}}\geq\mathbf{0}$, $\{\gamma^{k}\}>0$,
$\mathbf{x}^{0}\in X$, $\rho \in (0,1]$.

\hspace{1.24cm}Set $k=0$.

\texttt{$\mbox{(S.1)}:$}$\,\,$If $\mathbf{x}^{k}$ satisfies a termination
criterion: STOP;

\texttt{$\mbox{(S.2)}:$} For all $i\in\mathcal{N}$, solve (\ref{eq:decoupled_problem_i})
with accuracy $\varepsilon_{i}^{k}:$

\hspace{1.24cm}Find $\mathbf{z}_{i}^{k}\in X_i$ s.t. $\|\mathbf{z}_{i}^{k}-\widehat{\mathbf{x}}_i\left(\mathbf{x}^{k},\boldsymbol{{\tau}}\right)\|\leq\varepsilon_{i}^{k}$;

\texttt{$\mbox{(S.3)}:$}  Set $M^k \triangleq  \max_i \{E_i(\x^k)\}$.

\hspace{1.24cm}Choose a set $S^k$  that  contains at least one index $i$

\hspace{1.24cm}for which
$E_i(\x^k) \geq \rho M^k.$

\hspace{1.24cm}Set $\widehat{\z}^k_i =  \mathbf{\z}_{i}^k$ for $i\in S^k$ and
$\widehat{\z}^k_i = \x^k_i$ for $i\not \in S^k$

\texttt{$\mbox{(S.4)}:$} Set $\mathbf{x}^{k+1}\triangleq\mathbf{x}^k+\gamma^{k}\,(\widehat{\mathbf{z}}^{k}-\mathbf{x}^{k})$;

\texttt{$\mbox{(S.5)}:$} $k\leftarrow k+1$, and go to \texttt{$\mbox{(S.1)}.$}
\label{alg:general}
 \end{algo}\vspace{-0.4cm}

\begin{theorem} \label{Theorem_convergence_inexact_Jacobi}Let
$\{\mathbf{x}^{k}\}$ be the sequence generated by
Algorithm \ref{alg:general}, under A1-A6.
 Suppose  that $\{\gamma^{k}\}$
and $\{\varepsilon_{i}^{k}\}$ satisfy the following conditions: i)
$\gamma^{k}\in(0,1]$; ii) $\gamma^{k}\rightarrow0$; iii) $\sum_{k}\gamma^{k}=+\infty$;
iv) $\sum_{k}\left(\gamma^{k}\right)^{2}<+\infty$;
and v)  $\varepsilon_{i}^{k}\leq \gamma^k  \alpha_1\min\{\alpha_2, 1/\|\nabla_{\mathbf{x}_i} F(\mathbf{x}^k)\| \}$
for all $i\in {\cal N}$ and  some nonnegative  constants $\alpha_1$ and $\alpha_2$.
Additionally, if inexact solutions are used in Step S.2, i.e.,  $\varepsilon_{i}^{k}>0$ for some $i$ and infinite $k$, then 
assume also that $G$ is globally Lipschitz on $X$.

Then, either Algorithm \ref{alg:general} converges in a finite number of iterations to a stationary solution
of \eqref{eq:problem 1} or every limit point of
  $\{\mathbf{x}^{k}\}$ (at least
one such points exists) is a stationary solution of \eqref{eq:problem 1}.\vspace{-0.1cm}
\end{theorem}
In the theorem we obtain convergence to  stationary points $\x^*$, i.e.  points  for which  a subgradient $\xi \in \partial G(\x^*)$ exists such that
$(\nabla F(\x^*) $ $+ \xi)^T(\y-\x^*) \geq 0$ for all $\y\in X$. Of course, if $F$ is convex, stationary points coincide with  global minimizers.
\smallskip

\noindent \textbf{On Algorithm 1.} The proposed algorithm is extremely flexible. We can always choose $S^k ={\cal N}$ resulting in the simultaneous  update of all the (block) variables (full Jacobi scheme); or, at the other extreme, one can update a single (block) variable per time, thus obtaining a Gauss-Southwell kind of method.
One can also compute inexact solutions (Step 2) while preserving convergence, provided that  the error term $\varepsilon_{i}^{k}$  and  the step-size $\gamma^{k}$\textquoteright{}s are  chosen according
to Theorem 1. We emphasize that the Lipschitzianity of $G$ is required only if $\widehat{\mathbf{x}}(\mathbf{x}^k, \tau)$ is not computed exactly for infinite iterations. At any rate this Lipschitz conditions is automatically satisfied if $G$ is a norm (and therefore in Lasso and  group Lasso problems for example) or if
$X$ is bounded.\smallskip

\noindent \textbf{On the choice of the stepsize $\gamma^k$}\emph{.}
An example of step-size rule satisfying i-iv in Theorem \ref{Theorem_convergence_inexact_Jacobi}  is:
given $\gamma^{0}=1$, let
\begin{equation}\label{eq:gamma}
\gamma^{k}=\gamma^{k-1}\left(1-\theta\,\gamma^{k-1}\right),\quad k=1,\ldots,\vspace{-0.1cm}
\end{equation}
where $\theta\in(0,1)$ is a given constant; see \cite{scutari_facchinei_et_al_tsp13}
for others rules. This is actually the rule we used in our practical experiments, see next section. Notice that while this rule
may still require some tuning for optimal behaviour, it is quite reliable, since in general we are not
using a (sub)gradient direction, so that many of the well-known practical drawbacks associated
with a (sub)gradient method with diminishing step-size are mitigated in our setting. Furthermore, this choice of step-size does not require any form of centralized coordination, which is  a favourable feature in a parallel environment.

We remark that it is   possible to
prove convergence of Algorithm 1 also using other step-size rules, such as a standard Armijo-like line-search procedure or a (suitably small) constant step-size; see \cite{FacchineiSagratellaScutariMPsub13} for more details.
We omit the discussion of these  options because of lack of space, but also because
the former is not in line with our parallel approach while the latter is numerically less efficient.
\smallskip

\noindent \textbf{On the choice of $E_i(\mathbf{x})$}\emph{.}

\noindent $\bullet$
 As we mentioned, the most obvious choice is to take $E_i(\mathbf{x}) =  \|\widehat{\mathbf{x}}_{i}(\x^k,\tau_{i})-\mathbf{x}^k_i\|$.
This is a valuable choice if the computation of $\widehat{\mathbf{x}}_{i}(\x^k,\tau_{i})$ can be  easily accomplished. For instance, in the
Lasso problem with  ${\cal N} = \{1, \ldots, n\}$  (i.e., when each block reduces to a scalar variable),
it is well-known that  $\widehat{\mathbf{x}}_{i}(\x^k,\tau_{i})$ can be computed in closed form using
the soft-thresholding operator.

\noindent  $\bullet$ In situations where the computation of  $\|\widehat{\mathbf{x}}_{i}(\x^k,\tau_{i})-\mathbf{x}^k_i\|$ is not possible or advisable,
we can resort to estimates. To make the discussion simple, assume momentarily that $G\equiv 0$. Then it is known \cite{Facchinei-Pang_FVI03} that
$\|\Pi_{X_i} (\x^k_i - \nabla_{\x_i}F(\x^k)) - \x^k_i\|$  is an error bound for the minimization problem in \eqref{eq:decoupled_problem_i} and therefore
satisfies \eqref{eq:error bound}. In this situation we can choose $E_i(\x^k) = \|\Pi_{X_i} (\x^k_i - \nabla_{\x_i}F(\x^k)) - \x^k_i\|$.
If $G(\x) \not \equiv 0$ we can easily reduce to the case  $G\equiv 0$ by a simple transformation; the details are omitted for lack of space, see \cite{scutari_facchinei_et_al_tsp13}.

\noindent $\bullet$ It is interesting  to note that the computation of $E_i$ is only needed if a partial update of the (block) variables is
performed. However, an option that is always feasible is to take $S^k = {\cal N}$ at each iteration, i.e.,  update all (block) variables at each iteration.
With this choice we can  dispense with the computation of $E_i$ altogether.\smallskip

\noindent \textbf{On the choice of  $P_i(\mathbf{x}_i;\mathbf{x} )$}.

\noindent $\bullet$ The most obvious choice for $P_i$  is the linearization of $F$ at $\x^k$ with respect to $\x_i$:
$ P_i(\x_i; \x^k) = F(\x^k) + \nabla_{\x_i}F(\x^k)^T(\x_i - \x_i^k)$. With this choice, and taking for simplicity $\Q_i(\x^k) = \mathbf I$,
$\widehat {\mathbf x}_i(\x^k, \tau_i) $ is given by\vspace{-0.2cm}
\begin{equation}\label{eq:proposal 1}
 \underset{\x_{i}\in X_{i}}{\mbox{argmin}\,}  F(\x^k) + \nabla_{\x_i}F(\x^k)^T(\x_i - \x_i^k) + \frac{\tau_i}{2}
\| \x_i- \x_i^k\|^2 + g_i(\x_i).\vspace{-0.1cm}
\end{equation}
This is essentially the way   a new iteration is computed in most {\em sequential} (block-)CDMs for the
solution of (group) Lasso problems and its generalizations.  Note that contrary to most existing schemes,  our algorithm is {\em parallel}.

\noindent $\bullet$ At another extreme we could just take $P_i(\x_i; \x^k) = F(\x_i, \x_{-i}^k)$. Of course, to
have (P1) satisfied, we must assume that $F(\x_i, \x_{-i}^k)$ is convex.
 With this choice, and setting  for simplicity $\Q_i(\x^k) =\mathbf  I$, $\widehat \x_i(\x^k, \tau_i) $ is given by\vspace{-0.1cm}
\begin{equation}\label{eq:proposal 2}
 \underset{\x_{i}\in X_{i}}{\mbox{argmin}\,}  F(\x_i,  \x^k_{-i}) + \frac{\tau_i}{2}
\| \x_i- \x_i^k\|^2 + g_i(\x_i),
\end{equation}
thus giving rise to a parallel nonlinear Jacobi type method for the constrained minimization of $V(\x)$.

\noindent $\bullet$ Between the two ``extreme'' solutions proposed above one can consider ``intermediate'' choices. For example,
 If   $F(\x_i, \x_{-i}^k)$ is convex, we can take  $P_i(\x_i;\x^k)$ as a second order approximation of $F(\x_i, \x_{-i}^k)$, i.e.,
$ P_i(\x_i; \x^k) = F(\x^k) + \nabla_{\x_i}F(\x^k)^T$ $(\x_i - \x_i^k) + \frac{1}{2} (\x_i - \x_i^k)^T \nabla^2_{\x_i\x_i} F(\x^k) (\x_i - \x_i^k) $. When  $g_i(\x_i) \equiv 0$, this essentially corresponds to taking a Newton step in minimizing the ``reduced'' problem $\min_{\x_i\in X_i}F(\x_i, \x_{-i}^k)$. The resulting $\widehat \x_i(\x^k, \tau_i)$   is 
 \begin{align*}
 &\underset{\x_{i}\in X_{i}}{\mbox{argmin}\,}  F( \x^k) +\, \nabla_{\x_i}F( \x^k)^T(\x_i - \x_i^k) +
\\&\;
+\, \frac{1}{2} (\x_i -  \x_i^k)^T \nabla^2_{\x_i\x_i} F( \x^k) (\x_i - \x_i^k) +
 \frac{\tau_i}{2}\| \x_i- \x_i^k\|^2 + g_i(\x_i).
 \end{align*}\vspace{-0.4cm}

The framework described in Algorithm  \ref{alg:general}  can give rise
to very different instances, according to the choices one makes for the many
variable features it contains, some of which have been detailed above. For lack of space,  we cannot fully discuss here all possibilities.
We provide next   just a few instances  of possible algorithms that fall in our framework; more examples can be found in  \cite{FacchineiSagratellaScutariMPsub13}.

\noindent {\bf Example \#1$-$(Proximal) Jacobi algorithms for
convex functions: }
Consider the simplest problem falling in our setting: the unconstrained minimization of a continuously differentiable convex function, i.e.,
assume that $F$ is convex, $G(\x) \equiv 0$, and $X= \Re^n$. Although this is  possibly the best studied problem in nonlinear optimization, classical
parallel methods for this problem \cite[Sec. 3.2.4]{Bertsekas_Book-Parallel-Comp} require very strong contraction conditions. In our framework we can take $S^k = {\cal N}$,
$P_i(\x_i;\x^k) = F(\x_i, \x_{-i}^k)$, resulting in a fully parallel Jacobi-type  method which does not need any additional assumptions. Furthermore our theory shows that
we can even  dispense with the convexity assumption and still get convergence of a Jacobi-type method to a stationary point.

\noindent {\bf Example \# 2$-$Parallel coordinate descent method for Lasso}

\noindent
Consider the Lasso problem, i.e., $ F(\x) = \|\A\x -\bv\|^2 $,  $G(\x)= c \|\x\|_1$, and $X=\Re^n$. Probably, to date,  the most succesful class of methods for this problem is
that of CDMs, whereby at each iteration a single variable is updated  using \eqref{eq:proposal 1}.
We can easily obtain a parallel version for this method by taking $n_i =1$, $S^k = {\cal N}$ and
 still using  \eqref{eq:proposal 1}. Alternatively, instead of linearizing $F(\x)$, we can  better exploit the convexity of $F(\x)$ and  use   (\ref{eq:proposal 2}).  Furthermore, we can  easily consider similar methods for the group Lasso problem (just take $n_i >1$).
As a final remark, we observe  that convergence conditions of existing (deterministic) fully distributed parallel versions     of CDMs    such as
\cite{bradley2011parallel,yin2013parallel} impose a constraint on the maximum number of variables that can be simultaneously updated (linked to the  spectral radius of some matrices), a constraint that in many large scale problems is  likely not  satisfied. A key feature of the proposed scheme is that we can parallelize over (possibly) all variables while guaranteeing convergence.  \vspace{-0.3cm}

\section{Numerical Results{\small \vspace{-0.2cm}}}
In this section we report some preliminary numerical results that not only  show viability of our approach, but also seem to
indicate that our algorithmic framework can lead to practical methods that exploit well parallelism and compare favorably to existing schemes, both
parallel and sequential.
The tests were carried out on Lasso problems, the most studied case of Problem (\ref{eq:problem 1}) and, arguably, the most important one.
We generate four instances of problems using the  random generation technique proposed by Nesterov in \cite{nesterov2012gradient}, that permits to control the sparsity of
the solution. For the first three groups, we considered problems with 10,000 variables with the matrix $\A$ having 2,000 rows. The three groups differ in
the number of non zeros  of the solution, which is 20\%  (low sparsity), 10\% (medium sparsity), and 5\% (high sparsity) respectively.  The last group is
an instance   with 100,000 variables and  5000 rows, and  solutions having    5\% of non zero variables (high sparsity).

We implemented the instance of Algorithm 1  that we described in   Example \# 2 in the previous section, with the only difference that we used
\eqref{eq:proposal 2} instead of the proximal-linear choice \eqref{eq:proposal 1}.
Note that in the case of Lasso problems, the unique solution  \eqref{eq:proposal 2} can  be   computed
in closed form  using the soft-thresholding operator, see e.g. \cite{beck_teboulle_jis2009}. The free parameters of the algorithm are chosen as follows.
The proximal parameters are initially set to $\displaystyle \tau_i
= \text{tr}(\A^T\A)/2n$ for all $i$, where $n$ is the total number of variables. Furthermore,  we allowed a finite number of possible changes to $\tau_i$ according to the following rules:
(i) all $\tau_i$ are doubled  if at a certain iteration the objective function does not decrease; and (ii) they are all halved if the objective function decreases
for ten consecutive iterations. We updated $\gamma^k$ according to  \eqref{eq:gamma} with $\gamma^0 = 0.9$ and $\theta = 1e-5$. Note that since  $\tau_i$ are changed only a finite number of times,  conditions of  Theorem \ref{Theorem_convergence_inexact_Jacobi} are satisfied, and thus this instance of Algorithm 1 is guaranteed to converge.
Finally we  choose not to update all variables but set  $E_i(\x^k) =\|\widehat \x_i (\x^k,\tau_i) - \x_i^k\|$ and   $\rho = 0.5$ in Algorithm \ref{alg:general}. 

We compared our algorithm above,  termed FPA (for Flexible Parallel Algorithm), with a parallel implementation  of FISTA \cite{beck_teboulle_jis2009}, that can be regarded as the benchmark algorithm for Lasso problems, and Grock, a parallel algorithm proposed in \cite{yin2013parallel} that seems to perform extremely well on sparse problems. We actually tested two instances of Grock; in the first only one variable is updated at each iteration while in the second the number of updated variables is equal to the number of
parallel processors used (16 for the first three set of test problems, 32 for the last). Note that the theoretical convergence properties of Grock are in jeopardy as the number of updated variables increases and theoretical convergence conditions for this method are  likely to hold only if the columns of $\mathbf A$ are ``almost'' orthogonal, a feature enjoyed by our test problems, which however is not satisfied in most applications. As benchmark,  we also implemented two classical sequential schemes: (i) a Gauss-Seidel (GS) method computing  $\hat \x_i$, and
then updating  $\x_i$ using  unitary  step-size, in a sequential fashion, and (ii) a classical Alternating Method of Multipliers (ADMM)  \cite{boyd2011distributed} in the form of \cite{luo2012linear}. Note that ADMM can be parallelized, but they are known not to  scale well and therefore we did not consider this possibility.

\begin{figure}[h]
\vspace{-0.2cm}
\centering
        \begin{subfigure}[b]{0.25\textwidth}
                \includegraphics[width=\textwidth]{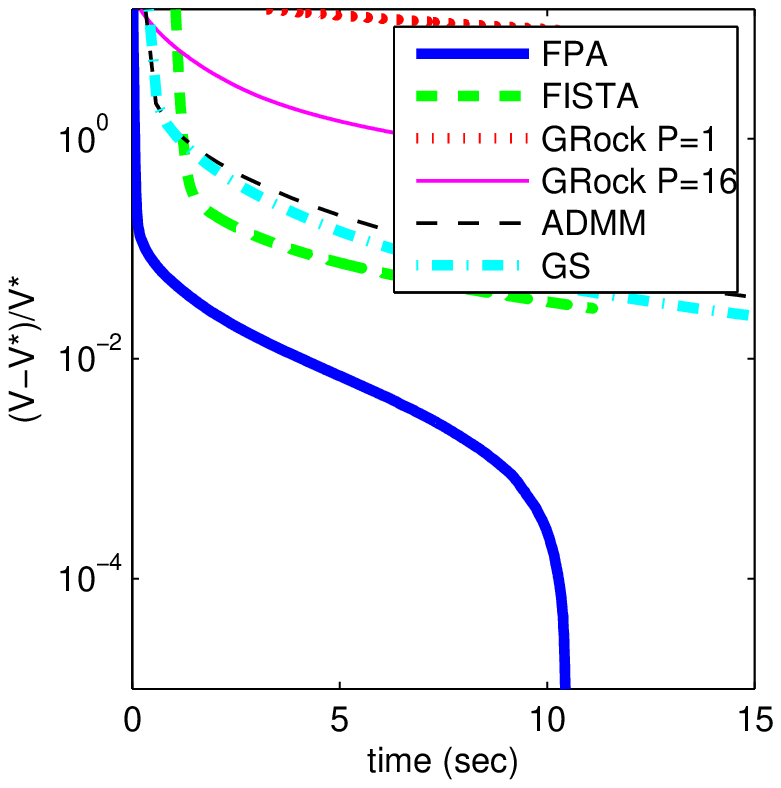}
                \caption{}
                \label{fig:low}
        \end{subfigure}
\hspace{-0.5cm}
        \begin{subfigure}[b]{0.25\textwidth}
\hspace{0.3cm}
                \includegraphics[width=\textwidth]{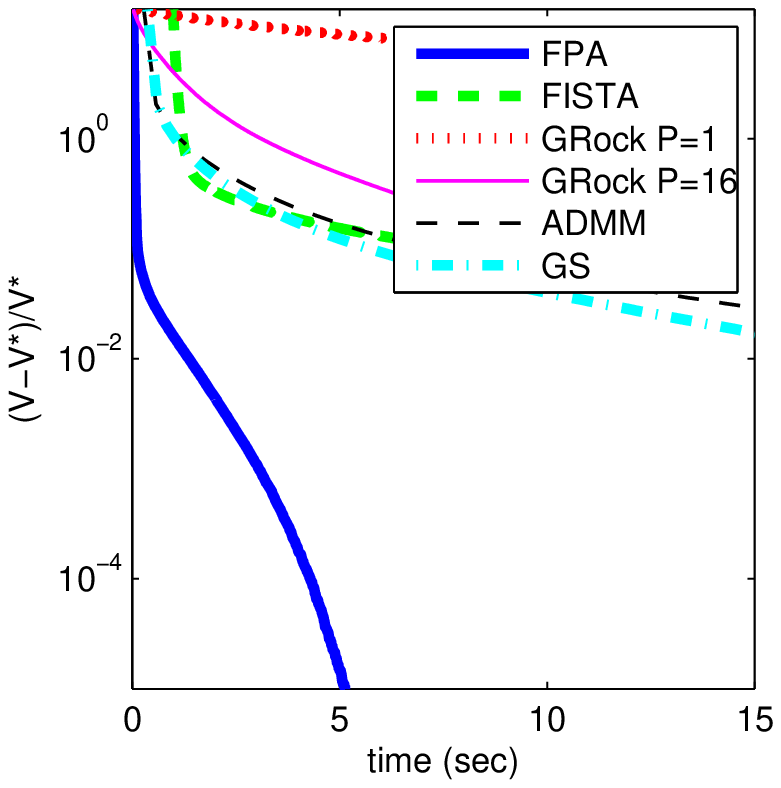}
                \caption{}
                \label{fig:med}
        \end{subfigure}
        \begin{subfigure}[b]{0.25\textwidth}
                \includegraphics[width=\textwidth]{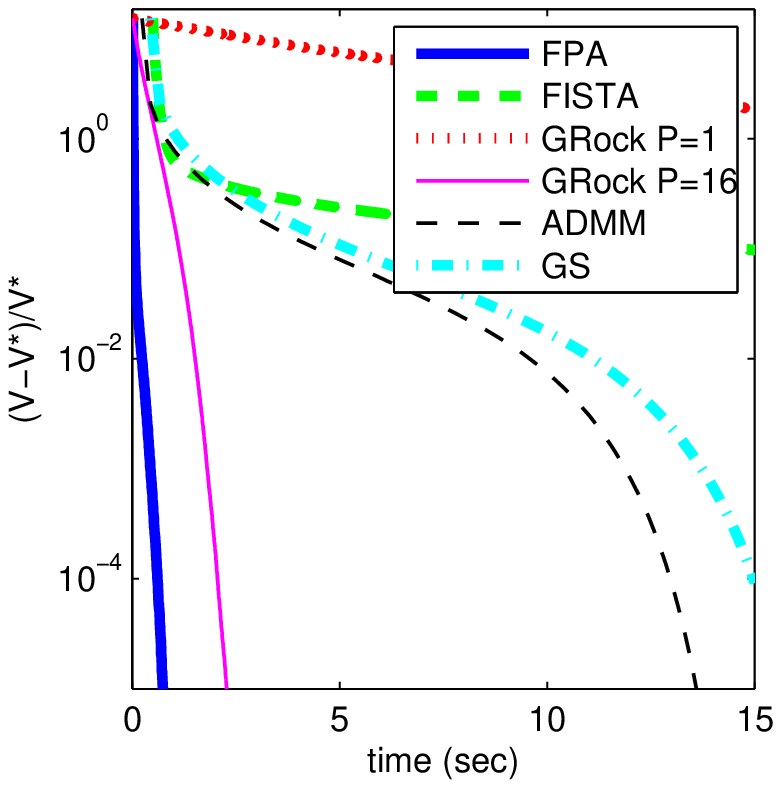}
                \caption{}
                \label{fig:high}
        \end{subfigure}
\hspace{-0.5cm}
        \begin{subfigure}[b]{0.25\textwidth}
\hspace{0.3cm}
                \includegraphics[width=\textwidth]{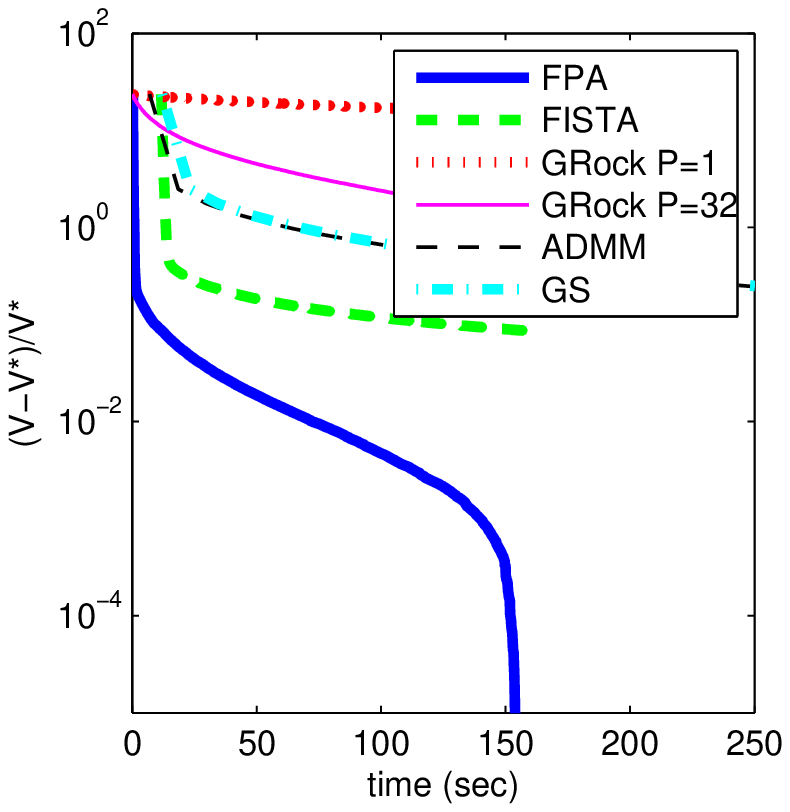}
                \caption{}
                \label{fig:big}
        \end{subfigure}\vspace{-0.4cm}
\caption{Relative error vs. time (in seconds, logarithmic scale):
(a) medium size and low sparsity - (b) medium size and  sparsity - (c) medium size and high sparsity - (d) large size and  high sparsity \label{fig}}\vspace{-0.3cm}
\end{figure}

All codes have been written in C++ and use the Message Passing Interface for parallel operations. All algebra is performed by using the GNU Scientific Library
(GSL). The algorithms were tested on a cluster at the State University of New York at Buffalo. All computations were done on one 32 core node composed  of four
8 core CPUs with 96GB of RAM  and Infiniband card. The 10,000 variables problems were solved using
$16$ parallel processes while for the 100,000 variables problems   $32$ parallel processes were used.  GS and ADMM were always run on a
single process. Results of our  experiments are reported  in Fig. \ref{fig}. The curves are  averaged over ten random realizations for each of the 10,000 variables groups, while for large 100,000 variables problems the average is  over 3 realizations.
 
Note that in Fig.\ref{fig} the CPU time includes communication times (for distributed algorithms) and the initial time needed by the methods to perform all pre-iterations computations (this explains why the  plot of FISTA starts  after the others; in fact FISTA requires some  nontrivial initializations based on the computation of  $\|\A\|_2^2$).

Some comments are in order. Fig \ref{fig} shows that on the tested problems FPA outperforms in a consistent manner all other implemented
algorithms.
Sequential methods behave strikingly  well on the 10,000 variables problems, if one keeps in mind that they only use one process; however, as expected,
they cannot compete with parallel methods when the dimensions increase. FISTA is capable to approach relatively fast  low accuracy  solutions, but has difficulties in
reaching high  accuracies. The parallel version of Grock is the closest match to FPA, but only when the problems are very sparse and the dimensions  not too large.
This is consistent
with the fact that at each iteration Grock updates only a very limited number of variables, and also with the fact that its convergence properties are at stake
when the problems are quite dense.   Our experiments also   suggest that, differently from what one could think (and often claimed in similar situations when using gradient-like methods),  updating only a (suitably chosen) subset of blocks rather than all variables may lead to  faster algorithms. 
In conclusion, we believe the results overall indicate that our approach can lead to very efficient practical methods for the solution of
large problems, with the flexibility to adapt to many different problem characteristics. 

\section{Conclusions}
We  proposed a highly parallelizable  algorithmic scheme for the minimization of the sum of a differentiable function and a block-separable nonsmooth one.
Our framework easily allows us to analyze parallel versions of well-known sequential methods and leads to entirely new algorithms.
When applied to large-scale Lasso problems, our algorithm was shown to outperform
existing methods.
\newpage

\section{Appendix: Proof of Theorem 1}
We first introduce some preliminary results instrumental to prove the theorem.  Hereafter, for   notational simplicity,  we will omit the dependence of $\widehat{\mathbf{x}}(\mathbf y,\boldsymbol{\tau})$ on $\boldsymbol\tau$ and write $\widehat{\mathbf{x}}(\mathbf y)$.  Given $S\subseteq \mathcal N$ and $\x\triangleq (x_i)_{i=1}^N$, we will also denote by $(\mathbf x)_S$ (or interchangeably $\mathbf x_S$) the vector whose component  $i$ is equal to $x_i$ if $i\in S$, and zero otherwise.  

\subsection{Intermediate results}

\begin{lemma}\label{Lemma_f_x_y_properties}  Set $H(\mathbf{x}; \mathbf{y}) \triangleq \sum_i h_i (\mathbf{x}_i; \mathbf{y})$. Then, the following hold:

\noindent (i) $H(\mathbf{\bullet};\mathbf{y})$
is uniformly strongly convex on $X$ with constant $c_{\boldsymbol{{\tau}}}>0$,
i.e., \vspace{-0.2cm}
\begin{equation}
\begin{array}{l}
\left(\mathbf{x}-\mathbf{w}\right)^{T}\left(\nabla_{\mathbf{x}}H\left(\mathbf{x};\mathbf{y}\right)-\nabla_{\mathbf{x}}H\left(\mathbf{w};\mathbf{y}\right)\right)\geq c_{{\boldsymbol{\tau}}}\left\Vert \mathbf{x}-\mathbf{w}\right\Vert ^{2}\end{array},\label{eq:strong_cvx_f_tilde}
\end{equation}
for all $\mathbf{x},\mathbf{w}\in X$ and given $\mathbf{y}\in X$;

\noindent  (ii) $\nabla_{\mathbf{x}}H(\mathbf{x};\mathbf{\bullet})$
is uniformly Lipschitz continuous on $X$, i.e., there exists
a  $0<L_{\nabla_H}<\infty$ independent on $\mathbf{x}$ such
that
\begin{equation}
\begin{array}{l}
\left\Vert \nabla_{\mathbf{x}}H\left(\mathbf{x};\mathbf{y}\right)-\nabla_{\mathbf{x}}H\left(\mathbf{x};\mathbf{w}\right)\right\Vert \end{array}\leq\, L_{\nabla H}\,\left\Vert \mathbf{y}-\mathbf{w}\right\Vert ,\label{eq:Lip_grad_L_f}
\end{equation}
for all $\mathbf{y},\mathbf{w}\in X$ and given $\mathbf{x}\in X$. \vspace{-0.3cm}\end{lemma}
\noindent  \textbf{Proof.} The proof is standard and thus is omitted.

\begin{proposition}\label{Prop_x_y} Consider  Problem (\ref{eq:problem 1})
under (A1)-(A6).
Then the mapping  $X\ni\mathbf{y}\mapsto\widehat{\mathbf{x}}(\mathbf{y})$
has the following properties:

\noindent (a)\emph{ $\widehat{\mathbf{x}} (\mathbf{\bullet})$}
is Lipschitz continuous on\emph{ $X$, }i.e., there
exists a positive constant $\hat{{L}}$ such that\emph{
\begin{equation}
\left\Vert \widehat{\mathbf{x}}(\mathbf{y})-\widehat{\mathbf{x}}(\mathbf{z})\right\Vert \leq\,\hat{{L}}\,\left\Vert \mathbf{y}-\mathbf{z}\right\Vert ,\quad\forall\mathbf{y},\mathbf{z}\in X;\vspace{-0.3cm}\label{eq:Lipt_x_map}
\end{equation}
}

\noindent(b) the set of the fixed-points of\emph{ }$\widehat{\mathbf{x}}(\mathbf{\bullet})$\emph{
}coincides with the set of stationary solutions of Problem
(\ref{eq:problem 1}); therefore $\widehat{\mathbf{x}}(\mathbf{y})$ has a fixed-point;

\noindent(c) for every given $\mathbf{y}\in X$   and for any set $S \subseteq {\cal N}$, it holds that
\begin{align} 
\left(\widehat{\mathbf{x}}(\mathbf{y})-\mathbf{y}\right)_S^{T}\,\nabla_{\mathbf{x}}F(\mathbf{y})_S + 
& \sum_{i\in S} g_i(\widehat{\mathbf{x}}_i (\mathbf{y})) - \sum_{i\in S} g_i(\mathbf{y}_i)\label{eq:descent_direction}
\\ &\leq-c_\tau\,\left\Vert (\widehat{\mathbf{x}}(\mathbf{y})-\mathbf{y})_S\right\Vert ^{2}, \nonumber \vspace{-0.3cm}
\end{align}
with $c_\tau \triangleq q\, \min_i \tau_i$.

%
\end{proposition}
\noindent \textbf{Proof.} We prove the proposition in the following order: (c), (a), (b).

\noindent (c): Given $\mathbf{y}\in X$, by definition,
each $\widehat{\mathbf{x}}_{i}(\mathbf{y})$ is the unique solution
of  problem (\ref{eq:decoupled_problem_i}); then it is not difficult to see that the following holds:
for all $\mathbf{z}_{i}\in X_{i}$,\vspace{-0.2cm}
\begin{equation}\label{eq:VI_i}
\left(  \mathbf{z}_i-\widehat{\mathbf{x}}_i(\mathbf{y})  \right)^T
\nabla_{\mathbf{x}_i}h_i(\widehat{\mathbf{x}}_{i}(\mathbf{y}); \mathbf{y}) + g_i(\mathbf{z}_i) - g_i(\widehat{\mathbf{x}}_{i}(\mathbf{y}))
\geq 0.
\end{equation}
Summing and subtracting $\nabla_{\mathbf{x}_{i}}P_i\left(\mathbf{y}_{i};\,\mathbf{y}\right)$
in (\ref{eq:VI_i}), choosing $\mathbf{z}_{i}=\mathbf{y}_{i}$, and
using  (P2),
we get
\begin{equation}
\begin{array}[t]{l}
\left(\mathbf{y}_{i}-\widehat{\mathbf{x}}_{i}(\mathbf{y})\right)^{T}\left(\nabla_{\mathbf{x}_{i}}P_{i}\!\left(\widehat{\mathbf{x}}_{i}(\mathbf{y});
\,\mathbf{y}\right)-\nabla_{\mathbf{x}_{i}}P_i\!\left(\mathbf{y}_{i};\,\mathbf{y}\right)\right)\smallskip\\
\quad+\left(\mathbf{y}_{i}-\widehat{\mathbf{x}}_{i}(\mathbf{y})\right)^{T}\nabla_{\mathbf{x}_{i}}F(\mathbf{y})\smallskip
 + g_i(\mathbf{y}_i) - g_i(\widehat{\mathbf{x}}_{i}(\mathbf{y}))
\\
\quad-\tau_{i}\,(\widehat{\mathbf{x}}_{i}(\mathbf{y})-\mathbf{y}_{i})^{T}\,\mathbf{Q}_{i}(\mathbf{y})\,(\widehat{\mathbf{x}}_{i}(\mathbf{y})-\mathbf{y}_{i})\geq0,
\end{array}\label{eq:VI_i_row2}
\end{equation}
for all $i\in\mathcal{N}$. Using (\ref{eq:VI_i_row2}) and observing that the term on the first line of
\eqref{eq:VI_i_row2} is non positive by (P1),
  we obtain \vspace{-0.1cm}
\begin{equation}
\left(\mathbf{y}_{i}-\widehat{\mathbf{x}}_{i}(\mathbf{y})\right)^{T}\nabla_{\mathbf{x}_{i}}F(\mathbf{y})
+ g_i(\mathbf{y}_i) - g_i(\widehat{\mathbf{x}}_{i}(\mathbf{y}))
\geq c_{\boldsymbol{{\tau}}}\left\Vert \widehat{\mathbf{x}}_{i}(\mathbf{y})-\mathbf{y}_{i}\right\Vert ^{2},\label{eq:VI_i_row4}
\end{equation}
for all $i\in\mathcal{N}$. Summing (\ref{eq:VI_i_row4}) over $i\in S$ 
we obtain (\ref{eq:descent_direction}).

\noindent(a): We use the notation introduced in Lemma \ref{Lemma_f_x_y_properties}.
Given $\mathbf{y},\mathbf{z}\in X$, by optimality and using \eqref{eq:VI_i}, we have
\[ \begin{array}{l}
\left(\mathbf{v}-\widehat{\mathbf{x}}(\mathbf{y})\right)^{T}\nabla_{\mathbf{x}}H\left(\widehat{\mathbf{x}}(\mathbf{y});\mathbf{y}\right)
 +G(\mathbf{v}) - G(\widehat{\mathbf{x}}(\mathbf{y}))
  \geq  0\;\forall\mathbf{v}\in X \\
\left(\mathbf{w}-\widehat{\mathbf{x}}(\mathbf{z})\right)^{T}\nabla_{\mathbf{x}}H\left(\widehat{\mathbf{x}}(\mathbf{z});\mathbf{z}\right)
 + G(\mathbf{w}) - G(\widehat{\mathbf{x}}(\mathbf{z})) \geq  0\;\forall\mathbf{w}\in X.
\end{array}
\]
 Setting $\mathbf{v}=\widehat{\mathbf{x}}(\mathbf{z})$ and $\mathbf{w}=\widehat{\mathbf{x}}(\mathbf{y})$,
summing the two inequalities above, and adding and subtracting $\nabla_{\mathbf{x}}H\left(\widehat{\mathbf{x}}(\mathbf{y});\mathbf{z}\right)$,
we obtain:
\begin{equation}
\begin{array}{l}
\left(\widehat{\mathbf{x}}(\mathbf{z})-\widehat{\mathbf{x}}(\mathbf{y})\right)^{T}\left(\nabla_{\mathbf{x}}H\left(\widehat{\mathbf{x}}(\mathbf{z});
\mathbf{z}\right)-\nabla_{\mathbf{x}}H\left(\widehat{\mathbf{x}}(\mathbf{y});\mathbf{z}\right)\right)\\
\leq\left(\widehat{\mathbf{x}}(\mathbf{y})-\widehat{\mathbf{x}}(\mathbf{z})\right)^{T}\left(\nabla_{\mathbf{x}}H\left(\widehat{\mathbf{x}}(\mathbf{y});
\mathbf{z}\right)-\nabla_{\mathbf{x}}H\left(\widehat{\mathbf{x}}(\mathbf{y});\mathbf{y}\right)\right).
\end{array}\label{eq:minimum_principle_Lip}
\end{equation}
 Using (\ref{eq:strong_cvx_f_tilde}) we can now lower bound the left-hand-side
of (\ref{eq:minimum_principle_Lip}) as
\begin{equation}
\begin{array}{l}
\left(\widehat{\mathbf{x}}(\mathbf{z})-\widehat{\mathbf{x}}(\mathbf{y})\right)^{T}\left(\nabla_{\mathbf{x}}H\left(\widehat{\mathbf{x}}(\mathbf{z});\mathbf{z}\right)-
\nabla_{\mathbf{x}}H\left(\widehat{\mathbf{x}}(\mathbf{y});\mathbf{z}\right)\right)\\
\,\,\geq c_{{\boldsymbol{\tau}}}\left\Vert \widehat{\mathbf{x}}(\mathbf{z})-\widehat{\mathbf{x}}(\mathbf{y})\right\Vert ^{2},
\end{array}\label{eq:lipschtz_map_2}
\end{equation}
whereas the right-hand side of (\ref{eq:minimum_principle_Lip}) can
be upper bounded as
\begin{equation}
\begin{array}{l}
\left(\widehat{\mathbf{x}}(\mathbf{y})-\widehat{\mathbf{x}}(\mathbf{z})\right)^{T}\left(\nabla_{\mathbf{x}}H\left(\widehat{\mathbf{x}}(\mathbf{y});\mathbf{z}\right)
-\nabla_{\mathbf{x}}H\left(\widehat{\mathbf{x}}(\mathbf{y});\mathbf{y}\right)\right)\\
\,\,\leq\, L_{\nabla H}\,\left\Vert \widehat{\mathbf{x}}(\mathbf{y})-\widehat{\mathbf{x}}(\mathbf{z})\right\Vert \,\left\Vert \mathbf{y}-\mathbf{z}\right\Vert ,
\end{array}\label{eq:lipschtz_map_1}
\end{equation}
where the inequality follows from the Cauchy-Schwartz inequality and
(\ref{eq:Lip_grad_L_f}). Combining (\ref{eq:minimum_principle_Lip}),
(\ref{eq:lipschtz_map_2}), and (\ref{eq:lipschtz_map_1}), we obtain
the desired Lipschitz property of $\widehat{\mathbf{x}}(\bullet)$.

\noindent (b): Let $\mathbf{x}^{\star}\in X$ be a fixed
point of $\widehat{\mathbf{x}}(\mathbf{y})$, that is $\mathbf{x}^{\star}=\widehat{\mathbf{x}}(\mathbf{x}^{\star})$.
Each $\widehat{\mathbf{x}}_{i}(\mathbf{y})$ satisfies (\ref{eq:VI_i}) for any given $\mathbf{y}\in X$.
For some $\boldsymbol\xi_i \in \partial g_i(\mathbf{x}^*)$, setting
$\mathbf{y}=\mathbf{x}^{\star}$ and using $\mathbf{x}^{\star}=\widehat{\mathbf{x}}(\mathbf{x}^{\star})$ and the convexity of $g_i$,
(\ref{eq:VI_i}) reduces to
\begin{equation}
\left(\mathbf{z}_{i}-\mathbf{x}_{i}^{\star}\right)^{T}(\nabla_{\mathbf{x}_{i}}F(\mathbf{x}^{\star}) + \boldsymbol\xi_i) \geq 0,\label{eq:fixed_point_min_principle}
\end{equation}
for all $\mathbf{z}_{i}\in X_{i}$ and $i\in\mathcal{N}$.
Taking into account the Cartesian structure of $X$, the separability of $G$,  and
summing (\ref{eq:fixed_point_min_principle}) over $i\in\mathcal{N}$
we obtain $\begin{array}[t]{l}
\left(\mathbf{z}-\mathbf{x}^{\star}\right)^{T}(\nabla_{\mathbf{x}}F(\mathbf{x}^{\star}) + \boldsymbol\xi) \geq0,\end{array}$ for all $\mathbf{z}\in X,$ with $\mathbf{z}\triangleq(\mathbf{z}_{i})_{i=1}^{N}$ and $\boldsymbol\xi \triangleq(\boldsymbol\xi_i)_{i=1}^{N} \in \partial G(\mathbf{x}^*)$;
therefore $\mathbf{x}^{\star}$ is a stationary solution of (\ref{eq:problem 1}).

The converse holds because i) $\widehat{\mathbf{x}}(\mathbf{x}^{\star})$
is the unique optimal solution of (\ref{eq:decoupled_problem_i})
with $\mathbf{y}=\mathbf{x}^{\star}$, and ii) $\mathbf{x}^{\star}$
is also an optimal solution of (\ref{eq:decoupled_problem_i}), since
it satisfies the minimum principle.\hfill $\square$

\begin{lemma} \emph{\cite[Lemma 3.4, p.121]{Bertsekas-Tsitsiklis_bookNeuro11}}\label{lemma_Robbinson_Siegmunt}
Let $\{X^{k}\}$, $\{Y^{k}\}$, and $\{Z^{k}\}$ be three sequences
of numbers such that $Y^{k}\geq0$ for all $k$. Suppose that
\[
X^{k+1}\leq X^{k}-Y^{k}+Z^{k},\quad\forall k=0,1,\ldots
\]
and $\sum_{k=0}^\infty Z^{k}<\infty$. Then either $X^{k}\rightarrow-\infty$
or else $\{X^{k}\}$ converges to a finite value and $\sum_{k=0}^\infty Y^{k}<\infty$.
\hfill $\Box$\end{lemma}

\begin{lemma}\label{lemma on errors} Let $\{\mathbf x^k\}$ be the sequence generated by Algorithm 1. Then, there is a positive constant $\tilde c$ such that the following holds: for all $k\geq 1$,
\begin{align*}
\left (\nabla_{\mathbf{x}}F (\mathbf{\x}^{k})\right)^{T}_{\tiny {S^k}} \left(\widehat{x}(\x^k)-\x^k\right)_{\tiny {S^k}}
+\sum_{i \in S^k} g_i(\widehat{\x}_i(\x^k)) -\sum_{i \in S^k} g_i(\x_i^k)\\ \hfill
 \leq
-\tilde c \,\| \widehat{\x}(\x^k)-\x^k \|^2.
\end{align*}\vspace{-0.4cm}\end{lemma}
\noindent \textbf{Proof.} Let $j_k$ be an index in $S^k$ such that $E_{j_k}(\x^k) \geq \rho \max_i E_i(\x^k)$ (Step 3 of the algorithm). Then, using the aforementioned bound and (\ref{eq:error bound}), 
it is easy to check that the following chain of inequalities holds:
\begin{align*}
\bar s_{j_k} \| \widehat{\x}_{S^k}(\x^k)  - \x^k_{S^k}\| & \geq \bar s_{j_k} \| \widehat{\x}_{j_k}(\x^k)  - \x^k_{j_k}\| \\ & \geq E_{j_k}(\x^k) \\ &
\geq \rho \max_i E_i(\x^k) \\ & \geq \left( \rho \min_i \underbar s_i\right) \left( \max_i \{ \|\widehat{\x}_i(\x^k)  - \x^k_i\| \} \right) \\ &
\geq \left( \frac{\rho \min_i \underbar s_i}{N}\right) \|\widehat{\x}(\x^k)  - \x^k\|.
\end{align*}
Hence we have for any $k$,
\begin{equation}\label{lower_bound_error_S_k}
\| \widehat{\x}_{S^k}(\x^k)  - \x^k_{S^k}\| \geq \left(\frac{\rho \min_i \underbar s_i}{N \bar s_{j_k}}\right) \|\widehat{\x}(\x^k)  - \x^k\|.
\end{equation}
Invoking now \ref{Prop_x_y} (c) with $S=S^k$ and $\y=\x^k$, and using \eqref{lower_bound_error_S_k}, the lemma holds, with $\tilde c \triangleq c_\tau \left(\frac{\rho \min_i \underbar s_i}{N \max_j \bar s_j}\right)^2$. \hfill $\Box$

\subsection{Proof of Theorem \ref{Theorem_convergence_inexact_Jacobi}} We are now ready to prove the theorem.
For any given $k\geq 0$, the Descent Lemma \cite{Bertsekas_NLPbook99}
yields
{\color{black}
\begin{equation}
\begin{array}{lll}
F\left(\mathbf{x}^{k+1}\right) & \leq & F\left(\mathbf{x}^{k}\right)+\gamma^{k}\,\nabla_{\mathbf{x}}F\left(\mathbf{x}^{k}\right)^{T}\left(\widehat{\mathbf{z}}^{k}-\mathbf{x}^{k}\right)\smallskip\\
 &  & +\dfrac{\left(\gamma^{k}\right)^{2}{L_{\nabla F}}}{2}\,\left\Vert \widehat{\mathbf{z}}^{k}-\mathbf{x}^{k}\right\Vert ^{2},
\end{array}\label{eq:descent_Lemma}
\end{equation}
with $\widehat{\mathbf{z}}^{k}\triangleq(\widehat{\mathbf{z}}_{i}^{k})_{i=1}^{N}$ and
$\mathbf{z}^{k}\triangleq(\mathbf{z}_{i}^{k})_{i=1}^{N}$ defined in Step 3 and 4  (Algorithm \ref{alg:general}).
Observe that
\begin{equation}\label{eq:49bis} \begin{array}{rcl}
 \left\Vert \widehat{\mathbf{z}}^{k}-\mathbf{x}^{k}\right\Vert ^{2}&\leq&
 \left\Vert \mathbf{z}^{k}-\mathbf{x}^{k}\right\Vert ^{2}\\
 & \leq &
 2\left\Vert \widehat{\mathbf{x}}(\mathbf{x}^{k})-\mathbf{x}^{k}\right\Vert ^{2}+2\sum_{i\in \cal N}\left\Vert \mathbf{z}_{i}^{k}-\widehat{\mathbf{x}}_{i}(\mathbf{x}^{k})\right\Vert ^{2}\\
& \leq& 2\left\Vert \widehat{\mathbf{x}}(\mathbf{x}^{k})-\mathbf{x}^{k}\right\Vert ^{2}+2\sum_{i \in \cal N}(\varepsilon_{i}^{k})^{2},
 \end{array}
 \end{equation}
 where the first inequality follows from the definition of $ \mathbf{z}^{k}$ and $ \widehat{\mathbf{z}}^{k}$ and
 in the last inequality we used $\left\Vert \mathbf{z}_{i}^{k}-\widehat{\mathbf{x}}_{i}(\mathbf{x}^{k})\right\Vert \leq\varepsilon_{i}^{k}$.

Denoting by $\overline{S}^k$ the complement of $S$, we also have, for $k$ large enough,
\begin{equation}
\begin{array}{l}
\nabla_{\mathbf{x}}F\left(\mathbf{x}^{k}\right)^{T}\left(\widehat{\mathbf{z}}^{k}-\mathbf{x}^{k}\right)\smallskip\\
\qquad\qquad   =
\nabla_{\mathbf{x}}F\left(\mathbf{x}^{k}\right)^{T}\left(\widehat{\mathbf{z}}^{k}-\widehat{\mathbf{x}}(\mathbf{x}^{k})
+\widehat{\mathbf{x}}(\mathbf{x}^{k})-\mathbf{x}^{k}\right)\smallskip\\
\qquad\qquad  = \nabla_{\mathbf{x}}F\left(\mathbf{x}^{k}\right)^{T}_{S^k} (\mathbf{z}^k - \widehat{\mathbf{x}}(\mathbf{x}^k))_{S^k}\smallskip\\ \qquad\qquad\quad
 +
 \nabla_{\mathbf{x}}F\left(\mathbf{x}^{k}\right)^{T}_{\overline{S}^k} (\mathbf{x}^k - \widehat{\mathbf{x}}(\mathbf{x}^k))_{\overline{S}^k}\smallskip\\
\qquad \qquad \quad   +  \nabla_{\mathbf{x}}F\left(\mathbf{x}^{k}\right)^{T}_{S^k} (\widehat{\mathbf{x}}(\mathbf{x}^k)-\mathbf{x}^k)_{S^k}\smallskip\\ \qquad\qquad\quad
 +    \nabla_{\mathbf{x}}F\left(\mathbf{x}^{k}\right)^{T}_{\overline{S}^k} (\widehat{\mathbf{x}}(\mathbf{x}^k)-\mathbf{x}^k)_{\overline{S}^k}\smallskip\\
\qquad\qquad   = \nabla_{\mathbf{x}}F\left(\mathbf{x}^{k}\right)^{T}_{S^k} (\mathbf{z}^k - \widehat{\mathbf{x}}(\mathbf{x}^k))_{S^k} \smallskip\\ \qquad\qquad +
 \nabla_{\mathbf{x}}F\left(\mathbf{x}^{k}\right)^{T}_{S^k} (\widehat{\mathbf{x}}(\mathbf{x}^k)-\mathbf{x}^k)_{S^k},
\end{array}
\end{equation}
where in the second equality we used the definition of $\widehat{\mathbf{z}}^k$ and of the set $S^k$.
Now, using the above identity and 
Lemma \ref{lemma on errors}, we can write

\begin{equation}\label{eq:descent_at_x_n}
\begin{array}{l}
\nabla_{\mathbf{x}}F\left(\mathbf{x}^{k}\right)^{T}\left(\widehat{\mathbf{z}}^{k}-\mathbf{x}^{k}\right)  + \sum_{i\in S^k} g_i(\widehat{\mathbf{z}}_i^{k})-\sum_{i\in S^k} g_i(\mathbf{x}^{k}_i)  \\[0.3em]
= \nabla_{\mathbf{x}}F\left(\mathbf{x}^{k}\right)^{T}\left(\widehat{\mathbf{z}}^{k}-\mathbf{x}^{k}\right)  +
\sum_{i\in S^k} g_i(\widehat{\mathbf{x}}_i(\mathbf{x}^{k})) -\sum_{i\in S^k} g_i(\mathbf{x}^{k}_i)  \\[0.3em]
\qquad \qquad +
 \sum_{i \in S^k} g_i(\widehat{\mathbf{z}}_i^k) -\sum_{i \in S^k} g_i(\widehat{\mathbf{x}}_i(\mathbf{x}^{k})) 
\\[0.3em]
\leq -\tilde c\left\Vert \widehat{\mathbf{x}}(\mathbf{x}^{k})-\mathbf{x}^{k}\right
\Vert ^{2}+\sum_{i \in S^k}\varepsilon_{i}^{k}\left\Vert \nabla_{\mathbf{x}_{i}}F(\mathbf{x}^{k})\right\Vert
+ L_G\sum_{i \in S^k}\varepsilon_{i}^{k}.
\end{array}
\end{equation}

Finally,  from    the definition of
 $\widehat{\z}^k$ and of the set $S^k$,  we have for all $k$ large enough,
\begin{equation}
\begin{array}{l}
V(\mathbf{x}^{k+1})  =  F(\mathbf{x}^{k+1}) + \sum_{ {i \in \cal N}}g_i (\mathbf{x}_i^{k+1}) \\[0.3em]
 =    F(\mathbf{x}^{k+1}) + \sum_{i\in {\cal N}}g_i (\mathbf{x}_i^{k} + \gamma^k(\widehat{\mathbf{z}}^k_i - \mathbf{x}_i^{k} ))\\[0.3em]
\leq   F(\mathbf{x}^{k+1}) +  \sum_{i \in \cal N}g_i (\mathbf{x}^k_i) + \gamma^k \left(\sum_{i \in S^k} (g_i (\widehat{\mathbf{z}}^k_i)
- g_i (\mathbf{x}^k_i)) \right)\\[0.3em]
\leq
V\left(\mathbf{x}^{k}\right)-\gamma^{k}\left(\tilde c -\gamma^{k}{L_{\nabla U}}\right)\left\Vert \widehat{\mathbf{x}}(\mathbf{x}^{k})-\mathbf{x}^{k}\right\Vert ^{2}+T^{k},\end{array}\label{eq:descent_Lemma_2}
\end{equation}
where in the first inequality we used the the convexity of the $g_i$'s, whereas the second follows from
 \eqref{eq:descent_Lemma}, \eqref{eq:49bis} and \eqref{eq:descent_at_x_n},  with    
$$T^{k}\triangleq\gamma^{k}\,
\sum_{i \in S^k}\varepsilon_{i}^{k}\left( L_G +
\left\Vert \nabla_{\mathbf{x}_{i}}F(\mathbf{x}^{k})\right\Vert\right) +\left(\gamma^{k}\right)^{2}{L_{\nabla F}}\,\sum_{i\in \cal N}(\varepsilon_{i}^{k})^{2}.$$
 
Using  assumption (v),  
we can bound $T^k$ as
\[
T^{k}\leq 
(\gamma^k)^2 \left[ N \alpha_1 (\alpha_2  L_G +1) + (\gamma^k)^2 L_{\nabla F} \left(N\alpha_1\alpha_2\right)^2 \right],
\]
 which, by assumption (iv) implies
 $\sum_{k=0}^{\infty}T^{k}<\infty$.
Since $\gamma^{k}\rightarrow 0$, it follows from (\ref{eq:descent_Lemma_2}) that there exist  some positive constant
$\beta_{1}$ and a sufficiently large $k$, say $k\geq\bar{{k}}$, such that
\begin{equation}
V(\mathbf{x}^{k+1})\leq V(\mathbf{x}^{k})-\gamma^{k}\beta_{1}\left\Vert \widehat{\mathbf{x}}(\mathbf{x}^{k})-\mathbf{x}^{k}\right\Vert ^{2}+T^{k}.\label{eq:descent_Lemma_3_}
\end{equation}
Invoking Lemma \ref{lemma_Robbinson_Siegmunt} with the identifications
$X^{k}=V(\mathbf{x}^{k+1})$, $Y^{k}=\gamma^{k}\beta_{1}\left\Vert \widehat{\mathbf{x}}(\mathbf{x}^{k})-\mathbf{x}^{k}\right\Vert ^{2}$
and $Z^{k}=T^{k}$ while using $\sum_{k=0}^\infty T^{k}<\infty$, we deduce
from (\ref{eq:descent_Lemma_3_}) that either $\{V(\mathbf{x}^{k})\}\rightarrow-\infty$
or else $\{V(\mathbf{x}^{k})\}$ converges to a finite
value and\vspace{-0.1cm}
\begin{equation}
\lim_{k\rightarrow\infty}\sum_{t=\bar{{k}}}^{k}\gamma^{t}\left\Vert \widehat{\mathbf{x}}(\mathbf{x}^{t})-\mathbf{x}^{t}\right\Vert ^{2}<+\infty.\vspace{-0.1cm}\label{eq:finite_sum_series}
\end{equation}
Since $V$ is coercive, $V(\mathbf{x})\geq\min_{\mathbf{y}\in X}V(\mathbf{y})>-\infty$,
implying that $\{V\left(\mathbf{x}^{k}\right)\}$ is convergent;
it follows from (\ref{eq:finite_sum_series}) and $\sum_{k=0}^{\infty}\gamma^{k}=\infty$
that $\liminf_{k\rightarrow\infty}\left\Vert \widehat{\mathbf{x}}(\mathbf{x}^{k})-\mathbf{x}^{k}\right\Vert =0.$

Using Prop. \ref{Prop_x_y}, we show next that $\lim_{k\rightarrow\infty}\left\Vert \widehat{\mathbf{x}}(\mathbf{x}^{k})-\mathbf{x}^{k}\right\Vert =0$;
for notational simplicity we will write $\triangle\widehat{\mathbf{x}}(\mathbf{x}^{k})\triangleq\widehat{\mathbf{x}}(\mathbf{x}^{k})-\mathbf{x}^{k}$.
Suppose, by contradiction, that $\limsup_{k\rightarrow\infty}\left\Vert \triangle\widehat{\mathbf{x}}(\mathbf{x}^{k})\right\Vert >0$.
Then, there exists a $\delta>0$ such that $\left\Vert \triangle\widehat{\mathbf{x}}(\mathbf{x}^{k})\right\Vert >2\delta$
for infinitely many $k$ and also $\left\Vert \triangle\widehat{\mathbf{x}}(\mathbf{x}^{k})\right\Vert <\delta$
for infinitely many $k$. Therefore, one can always find an infinite
set of indexes, say $\mathcal{K}$, having the following properties:
for any $k\in\mathcal{K}$, there exists an integer $i_{k}>k$ such
that  
\begin{eqnarray}
\left\Vert \triangle\widehat{\mathbf{x}}(\mathbf{x}^{k})\right\Vert <\delta, &  & \left\Vert \triangle\widehat{\mathbf{x}}(\mathbf{x}^{i_{k}})\right\Vert >2\delta\medskip\label{eq:outside_interval}\\
\delta\leq\left\Vert \triangle\widehat{\mathbf{x}}(\mathbf{x}^{j})\right\Vert \leq2\delta &  & k<j<i_{k}.\label{eq:inside_interval}
\end{eqnarray} 
Given the above bounds, the following holds: for all $k\in\mathcal{K}$,
\begin{eqnarray}
\delta & \overset{(a)}{<} & \left\Vert \triangle\widehat{\mathbf{x}}(\mathbf{x}^{i_{k}})\right\Vert -\left\Vert \triangle\widehat{\mathbf{x}}(\mathbf{x}^{k})\right\Vert \medskip\nonumber \\
 & \leq & \left\Vert \widehat{\mathbf{x}}(\mathbf{x}^{i_{k}})-\widehat{\mathbf{x}}(\mathbf{x}^{k})\right\Vert +\left\Vert \mathbf{x}^{i_{k}}-\mathbf{x}^{k}\right\Vert \\
 & \overset{(b)}{\leq} & (1+\hat{{L}})\left\Vert \mathbf{x}^{i_{k}}-\mathbf{x}^{k}\right\Vert \\
 & \overset{(c)}{\leq} &    {\color{black} (1+\hat{{L}})\sum_{t=k}^{i_{k}-1}\gamma^{t}\left(\left\Vert \triangle\widehat{\mathbf{x}}(\mathbf{x}^{t})_{S^t}\right\Vert +\left\Vert (\mathbf{z}^{t}-\widehat{\mathbf{x}}(\mathbf{x}^{t}))_{S^t}\right\Vert \right)}\vspace{-0.3cm}\nonumber \\
 & \overset{(d)}{\leq} & (1+\hat{{L}})\,(2\delta+\varepsilon^{\max})\sum_{t=k}^{i_{k}-1}\gamma^{t},\label{eq:lower_bound_sum}
\end{eqnarray}
where (a) follows from (\ref{eq:outside_interval});
(b) is due to Prop. \ref{Prop_x_y}(a); (c) comes from the triangle
inequality, the updating rule of the algorithm {\color{black} and the definition of $\widehat{\z}^k$}; and in (d) we used
(\ref{eq:outside_interval}), (\ref{eq:inside_interval}), and $\left\Vert \mathbf{z}^{t}-\widehat{\mathbf{x}}(\mathbf{x}^{t})\right\Vert \leq\sum_{i\in \cal N}\varepsilon_{i}^{t}$,
where $\varepsilon^{\max}\triangleq\max_{k}\sum_{i\in\cal N}\varepsilon_{i}^{k}<\infty$.
It follows from (\ref{eq:lower_bound_sum}) that
\begin{equation}
\liminf_{k\rightarrow\infty}\sum_{t=k}^{i_{k}-1}\gamma^{t}\geq\dfrac{{\delta}}{(1+\hat{{L}})(2\delta+\varepsilon^{\max})}>0.\label{eq:lim_inf_bound}
\end{equation}

\noindent
We show next that (\ref{eq:lim_inf_bound}) is in contradiction with
the convergence of $\{V(\mathbf{x}^{k})\}$. To do that, we preliminary
prove that, for sufficiently large $k\in\mathcal{K}$, it must be
$\left\Vert \triangle\widehat{\mathbf{x}}(\mathbf{x}^{k})\right\Vert \geq\delta/2$.
Proceeding as in (\ref{eq:lower_bound_sum}), we have: for any given
$k\in\mathcal{K}$,
\[
\begin{array}{l}
\left\Vert \triangle\widehat{\mathbf{x}}(\mathbf{x}^{k+1})\right\Vert -\left\Vert \triangle\widehat{\mathbf{x}}(\mathbf{x}^{k})\right\Vert \leq(1+\hat{{L}})\left\Vert \mathbf{x}^{k+1}-\mathbf{x}^{k}\right\Vert \smallskip\\
\qquad\qquad\qquad\qquad\qquad\leq(1+\hat{{L}})\gamma^{k}\left(\left\Vert \triangle\widehat{\mathbf{x}}(\mathbf{x}^{k})\right\Vert +\varepsilon^{\max}\right).
\end{array}
\]
 It turns out that for sufficiently large $k\in\mathcal{K}$ so that
$(1+\hat{{L}})\gamma^{k}<\delta/(\delta+2\varepsilon^{\max})$, it
must be
\begin{equation}
\left\Vert \triangle\widehat{\mathbf{x}}(\mathbf{x}^{k})\right\Vert \geq\delta/2;\label{eq:lower_bound_delta_x_n}
\end{equation}
otherwise the condition $\left\Vert \triangle\widehat{\mathbf{x}}(\mathbf{x}^{k+1})\right\Vert \geq\delta$
would be violated {[}cf. (\ref{eq:inside_interval}){]}. Hereafter
we assume w.l.o.g. that (\ref{eq:lower_bound_delta_x_n}) holds for
all $k\in\mathcal{K}$ (in fact, one can alway restrict $\{\mathbf{x}^{k}\}_{k\in\mathcal{K}}$
to a proper subsequence).

We can show now that (\ref{eq:lim_inf_bound}) is in contradiction
with the convergence of $\{V(\mathbf{x}^{k})\}$. Using (\ref{eq:descent_Lemma_3_})
(possibly over a subsequence), we have: for sufficiently large $k\in\mathcal{K}$,
\begin{eqnarray}
V(\mathbf{x}^{i_{k}}) & \leq & V(\mathbf{x}^{k})-\beta_{2}\sum_{t=k}^{i_{k}-1}\gamma^{t}\left\Vert \triangle\widehat{\mathbf{x}}(\mathbf{x}^{t})\right\Vert ^{2}+\sum_{t=k}^{i_{k}-1}T^{t}\nonumber \\
 & \overset{(a)}{<} & V(\mathbf{x}^{k})-\beta_{2}(\delta^{2}/4)\sum_{t=k}^{i_{k}-1}\gamma^{t}+\sum_{t=k}^{i_{k}-1}T^{t}\label{eq:liminf_zero}
\end{eqnarray}
where in (a) we used (\ref{eq:inside_interval}) and (\ref{eq:lower_bound_delta_x_n}),
and $\beta_{2}$ is some positive constant. Since $\{V(\mathbf{x}^{k})\}$
converges and $\sum_{k=0}^{\infty}T^{k}<\infty$, (\ref{eq:liminf_zero})
implies $\lim_{\mathcal{K}\ni k\rightarrow\infty}\,\sum_{t=k}^{i_{k}-1}\gamma^{t}=0,$
which contradicts (\ref{eq:lim_inf_bound}).

Finally, since the sequence $\{\mathbf{x}^{k}\}$ is bounded {[}due
to the coercivity of $V$ and the convergence of $\{V(\mathbf{x}^{k})\}${]},
it has at least one limit point $\bar{{\mathbf{x}}}$ that must belong
to $X$. By the continuity of $\widehat{\mathbf{x}}(\bullet)$
{[}Prop. \ref{Prop_x_y}(a){]} and $\lim_{k\rightarrow\infty}\left\Vert \widehat{\mathbf{x}}(\mathbf{x}^{k})-\mathbf{x}^{k}\right\Vert =0$,
it must be $\widehat{\mathbf{x}}(\bar{{\mathbf{x}}})=\bar{{\mathbf{x}}}$.
By Prop. \ref{Prop_x_y}(b) $\bar{{\mathbf{x}}}$ is also a stationary
solution of Problem  (\ref{eq:problem 1}).

 As a final remark, note that if $\varepsilon_i^k=0$ for every $i$ and for every $k$ large enough, i.e. if eventually
$\widehat{\mathbf{x}}(\mathbf{x}^k)$ is computed exactly, there is no need to assume that $G$ is
globally Lipschitz. In fact  in \eqref{eq:descent_at_x_n} the term containing $L_G$ disappears, and 
actually all the terms $T^k$ are zero and all the subsequent derivations independent of the Lipschitzianity of $G$.  
\hfill$\square$

\begin{spacing}{0.050000000000000003}
\bibliographystyle{IEEEtran}
\bibliography{scutari_facchinei_refs}

\begin{thebibliography}{10}
\providecommand{\url}[1]{#1}
\csname url@samestyle\endcsname
\providecommand{\newblock}{\relax}
\providecommand{\bibinfo}[2]{#2}
\providecommand{\BIBentrySTDinterwordspacing}{\spaceskip=0pt\relax}
\providecommand{\BIBentryALTinterwordstretchfactor}{4}
\providecommand{\BIBentryALTinterwordspacing}{\spaceskip=\fontdimen2\font plus
\BIBentryALTinterwordstretchfactor\fontdimen3\font minus
  \fontdimen4\font\relax}
\providecommand{\BIBforeignlanguage}[2]{{%
\expandafter\ifx\csname l@#1\endcsname\relax
\typeout{** WARNING: IEEEtran.bst: No hyphenation pattern has been}%
\typeout{** loaded for the language `#1'. Using the pattern for}%
\typeout{** the default language instead.}%
\else
\language=\csname l@#1\endcsname
\fi
#2}}
\providecommand{\BIBdecl}{\relax}
\BIBdecl

\bibitem{bach2011optimization}
F.~Bach, R.~Jenatton, J.~Mairal, and G.~Obozinski, ``{Optimization with
  sparsity-inducing penalties},'' \emph{arXiv preprint arXiv:1108.0775}, 2011.

\bibitem{bradley2011parallel}
J.~K. Bradley, A.~Kyrola, D.~Bickson, and C.~Guestrin, ``{Parallel coordinate
  descent for l1-regularized loss minimization},'' \emph{arXiv preprint
  arXiv:1105.5379}, 2011.

\bibitem{buhlmann2011statistics}
P.~L. B{\"u}hlmann, S.~A. van~de Geer, and S.~{Van de Geer}, \emph{{Statistics
  for high-dimensional data}}.\hskip 1em plus 0.5em minus 0.4em\relax Springer,
  2011.

\bibitem{byrd2013inexact}
R.~H. Byrd, J.~Nocedal, and F.~Oztoprak, ``{An Inexact Successive Quadratic
  Approximation Method for Convex L-1 Regularized Optimization},'' \emph{arXiv
  preprint arXiv:1309.3529}, 2013.

\bibitem{fountoulakis2013second}
K.~Fountoulakis and J.~Gondzio, ``{A Second-Order Method for Strongly Convex
  L1-Regularization Problems},'' \emph{arXiv preprint arXiv:1306.5386}, 2013.

\bibitem{necoara2013efficient}
I.~Necoara and D.~Clipici, ``{Efficient parallel coordinate descent algorithm
  for convex optimization problems with separable constraints: application to
  distributed MPC},'' \emph{Journal of Process Control}, vol.~23, no.~3, pp.
  243--253, 2013.

\bibitem{nesterov2012gradient}
Y.~Nesterov, ``{Gradient methods for minimizing composite functions},''
  \emph{Mathematical Programming}, pp. 1--37, 2012.

\bibitem{nesterov2012efficiency}
------, ``{Efficiency of coordinate descent methods on huge-scale optimization
  problems},'' \emph{SIAM Journal on Optimization}, vol.~22, no.~2, pp.
  341--362, 2012.

\bibitem{qin2010efficient}
Z.~Qin, K.~Scheinberg, and D.~Goldfarb, ``{Efficient block-coordinate descent
  algorithms for the group lasso},'' \emph{Mathematical Programming
  Computation}, pp. 1--27, 2010.

\bibitem{rakotomamonjy2011surveying}
A.~Rakotomamonjy, ``{Surveying and comparing simultaneous sparse approximation
  (or group-lasso) algorithms},'' \emph{Signal processing}, vol.~91, no.~7, pp.
  1505--1526, 2011.

\bibitem{razaviyayn2013unified}
M.~Razaviyayn, M.~Hong, and Z.-Q. Luo, ``{A unified convergence analysis of
  block successive minimization methods for nonsmooth optimization},''
  \emph{SIAM Journal on Optimization}, vol.~23, no.~2, pp. 1126--1153, 2013.

\bibitem{richtarik2012iteration}
P.~Richt{\'a}rik and M.~Tak{\'a}\v{c}, ``{Iteration complexity of randomized
  block-coordinate descent methods for minimizing a composite function},''
  \emph{Mathematical Programming}, pp. 1--38, 2012.

\bibitem{richtarik2012parallel}
------, ``{Parallel coordinate descent methods for big data optimization},''
  \emph{arXiv preprint arXiv:1212.0873}, 2012.

\bibitem{Sra-Nowozin-Wright_book11}
S.~Sra, S.~Nowozin, and S.~J. Wright, Eds., \emph{{Optimization for Machine
  Learning}}, ser. {Neural Information Processing}.\hskip 1em plus 0.5em minus
  0.4em\relax Cambridge, Massachusetts: The MIT Press, Sept. 2011.

\bibitem{tseng2009coordinate}
P.~Tseng and S.~Yun, ``{A coordinate gradient descent method for nonsmooth
  separable minimization},'' \emph{Mathematical Programming}, vol. 117, no.
  1-2, pp. 387--423, 2009.

\bibitem{xu2012block}
\BIBentryALTinterwordspacing
Y.~Xu and W.~Yin, ``{A block coordinate descent method for multi-convex
  optimization with applications to nonnegative tensor factorization and
  completion},'' DTIC Document, Tech. Rep., 2012. [Online]. Available:
  \url{http://www.caam.rice.edu/$\sim$optimization/bcu/multiconvex.html}
\BIBentrySTDinterwordspacing

\bibitem{yin2013parallel}
\BIBentryALTinterwordspacing
Z.~Yin, P.~Ming, and Y.~Wotao, ``{Parallel and Distributed Sparse
  Optimization},'' 2013. [Online]. Available:
  \url{http://www.caam.rice.edu/$\sim$optimization/disparse/}
\BIBentrySTDinterwordspacing

\bibitem{yuan2010comparison}
G.-X. Yuan, K.-W. Chang, C.-J. Hsieh, and C.-J. Lin, ``{A comparison of
  optimization methods and software for large-scale l1-regularized linear
  classification},'' \emph{The Journal of Machine Learning Research}, vol.
  9999, pp. 3183--3234, 2010.

\bibitem{wright2012accelerated}
S.~J. Wright, ``{Accelerated block-coordinate relaxation for regularized
  optimization},'' \emph{SIAM Journal on Optimization}, vol.~22, no.~1, pp.
  159--186, 2012.

\bibitem{scutari_facchinei_et_al_icassp13}
G.~Scutari, F.~Facchinei, P.~Song, D.~P. Palomar, and J.-S. Pang,
  ``{Decomposition by partial linearization in multiuser systems},'' in
  \emph{{IEEE International Conference on Acoustics, Speech and Signal
  Processing (ICASSP 2013)}}, May 4-9 2013, pp. 4424--4428.

\bibitem{scutari_facchinei_et_al_tsp13}
G.~Scutari, F.~Facchinei, P.~Song, D.~Palomar, and J.-S. Pang, ``{Decomposition
  by Partial linearization: Parallel optimization of multi-agent systems},''
  \emph{IEEE Transactions on Signal Processing}, to appear, 2013.

\bibitem{tibshirani1996regression}
R.~Tibshirani, ``{Regression shrinkage and selection via the lasso},''
  \emph{Journal of the Royal Statistical Society. Series B (Methodological)},
  pp. 267--288, 1996.

\bibitem{yuan2006model}
M.~Yuan and Y.~Lin, ``{Model selection and estimation in regression with
  grouped variables},'' \emph{Journal of the Royal Statistical Society: Series
  B (Statistical Methodology)}, vol.~68, no.~1, pp. 49--67, 2006.

\bibitem{shevade2003simple}
S.~K. Shevade and S.~S. Keerthi, ``{A simple and efficient algorithm for gene
  selection using sparse logistic regression},'' \emph{Bioinformatics},
  vol.~19, no.~17, pp. 2246--2253, 2003.

\bibitem{meier2008group}
L.~Meier, S.~{Van De Geer}, and P.~B{\"u}hlmann, ``{The group lasso for
  logistic regression},'' \emph{Journal of the Royal Statistical Society:
  Series B (Statistical Methodology)}, vol.~70, no.~1, pp. 53--71, 2008.

\bibitem{goldfarb2012fast}
D.~Goldfarb, S.~Ma, and K.~Scheinberg, ``{Fast alternating linearization
  methods for minimizing the sum of two convex functions},'' \emph{Mathematical
  Programming}, pp. 1--34, 2012.

\bibitem{Bertsekas_Book-Parallel-Comp}
D.~P. Bertsekas and J.~N. Tsitsiklis, \emph{{Parallel and Distributed
  Computation: Numerical Methods}}, 2nd~ed.\hskip 1em plus 0.5em minus
  0.4em\relax Athena Scientific Press, 1989.

\bibitem{FacchineiSagratellaScutariMPsub13}
\BIBentryALTinterwordspacing
F.~Facchinei, S.~Sagratella, and G.~Scutari, ``{Flexible Parallel Algorithms
  for Big Data Optimization},'' Dept. of Electrical Eng., State University of
  New York at Buffalo, Buffalo, NY, USA, Tech. Rep., 2013. [Online]. Available:
  \url{http://www.eng.buffalo.edu/$\sim$gesualdo/FSSTechRep13.pdf.}
\BIBentrySTDinterwordspacing

\bibitem{Facchinei-Pang_FVI03}
F.~Facchinei and J.-S. Pang, \emph{{Finite-Dimensional Variational Inequalities
  and Complementarity Problem}}.\hskip 1em plus 0.5em minus 0.4em\relax
  Springer-Verlag, New York, 2003.

\bibitem{beck_teboulle_jis2009}
A.~Beck and M.~Teboulle, ``{A fast iterative shrinkage-thresholding algorithm
  for linear inverse problems},'' \emph{SIAM Journal on Imaging Sciences},
  vol.~2, no.~1, pp. 183--202, 2009.

\bibitem{boyd2011distributed}
S.~Boyd, N.~Parikh, E.~Chu, B.~Peleato, and J.~Eckstein, ``{Distributed
  optimization and statistical learning via the alternating direction method of
  multipliers},'' \emph{Foundations and Trends{\textregistered} in Machine
  Learning}, vol.~3, no.~1, pp. 1--122, 2011.

\bibitem{luo2012linear}
Z.-Q. Luo and M.~Hong, ``{On the linear convergence of the alternating
  direction method of multipliers},'' \emph{arXiv preprint arXiv:1208.3922},
  2012.

\bibitem{Bertsekas-Tsitsiklis_bookNeuro11}
D.~P. Bertsekas and J.~N. Tsitsiklis, \emph{{Neuro-Dynamic Programming}}.\hskip
  1em plus 0.5em minus 0.4em\relax Cambridge, Massachusetts: Athena Scientific
  Press, May. 2011.

\bibitem{Bertsekas_NLPbook99}
D.~Bertsekas, \emph{{Nonlinear Programming}}.\hskip 1em plus 0.5em minus
  0.4em\relax Belmont, MA, USA: Athena Scientific, 2th Ed., 1999.

\end{thebibliography}
\end{spacing}

\end{document}